\documentstyle[preprint,revtex]{aps}
\begin{document}
\draft
\begin{title}
Chiral Bosons Through Linear Constraints
\end{title}
\author{H. O. Girotti \cite{ast}, M. Gomes and V. O. Rivelles}
\begin{instit}
Instituto de F\'{\i}sica, Universidade de S\~ao Paulo,\\
Caixa Postal 20516,  01498 S\~ao Paulo, SP, Brazil.
\end{instit}
\begin{abstract}
We study in detail the quantization of a model which apparently
describes chiral bosons. The model is based on the idea that the
chiral condition could be implemented through a linear constraint.
We show that the space of states is of indefinite metric. We cure
this disease by introducing ghost fields in such a way that a
BRST symmetry is generated. A quartet algebra is seen to emerge.
The quartet mechanism, then, forces all physical states, but the vacuum,
to have zero norm.
\end{abstract}
\narrowtext
It has been claimed in the literature \cite{Pr,Mu} that the two dimensional
Lorentz
invariant model \cite{um}
\begin{equation}
{\cal L} =\frac{1}{2} (\partial_\mu \varphi)  (\partial^\mu
\varphi)+\lambda_\mu (g^{\mu \nu}-\epsilon^{\mu \nu})\partial_\nu \varphi
\label{one}
\end{equation}
 describes  chiral bosons, namely, a field satisfying the
 equation $\partial_{-}\varphi \equiv (\partial_{\scriptstyle 0}
-\partial_1)\varphi=0$.  The procedure for constructing
 the Lagrangian (\ref{one})  is rather obvious, the chiral condition
 has been  ``linearly'' added
to the Lagrangian of a free massless
scalar field through the Lagrange multiplier
 $\lambda_\mu$. However,  from the  equations  of motion  deriving
from (\ref{one}),
\begin{mathletters}
\begin{eqnarray}
&&\partial_{\mu}\partial^{\mu} \varphi + (g^{\mu \nu} - \epsilon^{\mu \nu})
\partial_{\nu} \lambda_{\mu}=0, \\
&&(g^{\mu \nu} - \epsilon^{\mu \nu})
\partial_{\nu} \varphi=0, \\
&&(g^{\mu \nu} - \epsilon^{\mu \nu})
\partial_{\nu} \lambda_{\mu}=0,
\end{eqnarray}
\end{mathletters}
one sees that not only $\varphi$ but also $\lambda_\mu$ are chiral fields. The
fact that the Lagrange multiplier
$\lambda_\mu$ becomes dynamical  was first
noticed  by Siegel \cite{Si}.

Within the Hamiltonian formulation, the model is specified by the canonical
Hamiltonian
\begin{equation}
H_{0} = \int dx \Pi(x) \varphi^{\prime}(x), \label{H0}
\end{equation}
together with the second class constraints
\begin{mathletters}
\begin{eqnarray}
&& T_{1}(x) \equiv p_{\lambda}(x) \approx 0,  \label{T1} \\
&& T_{2}(x) \equiv \lambda_{+}(x) - \Pi(x) + \varphi^{\prime}(x)\approx 0,
\label{T2}
\end{eqnarray}
\end{mathletters}
where  $\Pi$ and $p_{\lambda}$ are the canonical conjugate momenta of
$\varphi$ and $\lambda_{+}
\equiv \lambda_0 +\lambda_1$, respectively. Furthermore, $\varphi^{\prime}$
($\dot \varphi$)
is a shorthand notation for $\partial_1\varphi$ ($\partial_0 \varphi$).
The above constraints   allow for the elimination of the sector
$\lambda_{+}$, $p_{\lambda}$ from the phase space.  The reduced phase space
is then spanned by the variables $\varphi$ and $\Pi$ whose Dirac
brackets \cite{Di} are, as they must  \cite{Fr}, equal to the
corresponding Poisson
brackets. Hence, when formally quantized according to the Dirac
bracket
procedure \cite{Di}, the theory appears to describe a single
chiral field \cite{Pr}.

In this paper we study in detail the particle content of the model. As we
shall see, the metric of the space of states is not positive definite.
We cure this problem by adding ghosts to the original Lagrangian so that
a BRST symmetry emerges. We  demonstrate, afterwards, that the original fields
 and the
ghosts obey a quartet algebra \cite{Ku}. Then,  the quartet mechanism
\cite{Ku},
when applied to this case,  leads to the conclusion that the only surviving
state of positive norm is the vacuum state. Thus the model is not appropriate
to describe neither chiral bosons nor any other quantum excitation.

As pointed out in ref.  \cite{Pr}, the quantum equations of motion
obeyed by the fields $\varphi$ and $\Pi $ are $\partial_{-} \varphi =0$
and $\partial_{-} \Pi = 0 $. These equations and the canonical equal--time
commutation relations are solved by ($x^{+}\equiv x^0+x^1$)
\begin{eqnarray}
&& \varphi(x^{+}) = \frac {1}{\sqrt {2\pi}} \int_{0}^{\infty}dp [{\rm
e}^{-ipx^+}
a(p)+{\rm e}^{ipx^+}a^{\dag}(p)], \\
&& \Pi(x^{+})=  \frac {1}{\sqrt {2\pi}} \int_{0}^{\infty}dp [{\rm e}^{-ipx^+}
b(p)+{\rm e}^{ipx^+}b^{\dag}(p)],
\end{eqnarray}
with
\begin{equation}
  [a(p),\, b^{\dag}(p^{\prime})]=- [b(p),\, a^{\dag}(p^{\prime})]  =
 i \delta(p-p^{\prime}) \label{two}
\end{equation}
as the only nonvanishing commutators.

The normal ordered quantum counterpart of  the classical Hamiltonian $H_0$ is

\begin{equation}
H_0= i\int_{0}^{\infty}dp \, p [a^{\dag} (p) b (p) - b^{\dag}(p) a(p)].
\end{equation}
To make explicit that the space of states we are dealing with is  of
indefinite metric, we introduce the operators
\begin{eqnarray}
A& \equiv & \frac {1}{\sqrt 2} (a+ i b) \\
B& \equiv & \frac {1}{\sqrt 2} (a- i b),
\end{eqnarray}
which are easily seen to obey the commutation relations
\begin{equation}
  [A(p),\, A^{\dag}(p^{\prime})]=- [B(p),\, B^{\dag}(p^{\prime})]  =
\delta (p - p^{\prime}).
\end{equation}
It is now clear that all  states
obtained by applying to the vacuum the operator $B^{\dag}$  an odd number of
times are of negative norm.
In terms of $A$ and $B$ the Hamiltonian assumes the standard form
\begin{equation}
H_0= \int_{0}^{\infty}dp \, p [A^{\dag} (p) A (p) - B^{\dag}(p) B(p)].
\end{equation}
The lack of boundedness of $H_{0}$ at the classical level reflects
itself, at the quantum level, through the appearance of states of
negative norm.

To cure the disease represented by the states of negative norm, we bring
into
the theory the real  Grassmann fields $\bar C_{\mu}(x)$
 and $C(x)$. This is done  by adding to ${\cal L}$ the ghost
Lagrangian
\begin{equation}
{\cal L}_g= i  C\partial_{-}\bar C_{+},
\end{equation}
where $\bar C_{+}\equiv \bar C_0 + \bar C_{1}$.
One can corroborate that ${\cal L}_T \equiv {\cal L}+ {\cal L}_g$ is invariant
under the global nilpotent  transformation
\begin{mathletters}
\label{10}
\begin{eqnarray}
\delta \varphi(x)& = &i \epsilon_{-} \bar C_{+}(x)\\
\delta \lambda_{+}(x) & = & i \epsilon_{-} \partial_{+}\bar C_{+}(x)\\
\delta \bar C_{+}(x) &=& 0\\
\delta C(x) & =& -i \epsilon_{-}\lambda_{+}(x)- \frac {i}{2}
\epsilon_{-}\partial_{+}
\varphi(x).
\end{eqnarray}
\end{mathletters}
We emphasize that the original bosonic Lagrangian does not posses a local
symmetry since it only exhibits second class constraints. Nevertheless, a
BRST symmetry has emerged after the addition of the ghost fields. An
analogous situation has already been encountered in the literature \cite{TK}.

The canonical ghost Hamiltonian
\begin{equation}
H_{0}^{g}=i\int dx C(x)\bar C_{+}^{\prime}(x)\label{four}
\end{equation}
and the second class constraints
\begin{mathletters}
\begin{eqnarray}
T_{1}^{g}(x)& \equiv & p(x)\approx 0, \label{three}\\
T_{2}^{g}(x)& \equiv & \bar p_{-}(x)-i C(x)\approx 0, \label{five}
\end{eqnarray}
\end{mathletters}
define the dynamics of the ghost fields in the Hamiltonian framework.
Here, $p$ and
$\bar p_{-}$ are the canonical conjugate momenta of $C$ and $\bar C_{+}$,
respectively. Clearly, the sector $ C $, $ p $ can be
be eliminated from
phase space, although, following common practice, we shall keep
$ \bar C_{+} $ and $ C $ as the canonical variables spanning
the ghost sector of the reduced phase space. As required
\cite{Fr}, the Dirac brackets involving $ \bar C_{+} $ and $ C $
equal the corresponding generalized Poisson brackets. The quantum
counterpart of $H_{0}^{g}$ is obtained from (\ref{four}) after
appropriate symmetrization, required to solve the ordering problem.
The equations of motion obeyed by the ghost field operators  are, then,
found to be $\partial_{-} C = 0 $ and $\partial_{-} \bar C_{+} = 0 $.
These equations and the canonical equal-time anticommutation relations
are solved by
\begin{eqnarray}
&& C (x^{+}) = \frac {1}{\sqrt {2\pi}} \int_{0}^{\infty}dp [{\rm e}^{-ipx^+}
d(p)+{\rm e}^{ipx^+}d^{\dag}(p)], \label{six}  \\
&& \bar C_{+}(x^{+})=  \frac {1}{\sqrt {2\pi}} \int_{0}^{\infty}dp [{\rm
e}^{-ipx^+}
\bar d(p)+{\rm e}^{ipx^+}{\bar d}^{\dag}(p)], \label{seven}
\end{eqnarray}
where the nonvanishing anticommutators are
\begin{equation}
\{ d(p), \bar d^{\dag}(p{^\prime}) \} = \{ d^{\dag}(p), \bar d(p{^\prime}) \}
= \delta (p - p^{\prime} ). \label{8}
\end{equation}
By replacing (\ref{six}) and (\ref{seven}) in  $ H_{0}^{g} $ one arrives to
\begin{equation}
 H_{0}^{g} = \int_{0}^{\infty}dp \, p\,[{\bar d}^{\dag}(p) d(p) + d^{\dag}(p)
\bar d(p)] ,
\end{equation}
where the normal ordering prescription has been used. After attributing ghost
number $-1$ and $+1$ to $ C $ and $ \bar C_{+} $, respectively, one finds that
\begin{equation}
iN_{g} = \int_{0}^{\infty} dp [{\bar d}^{\dag}(p) d(p) - d^{\dag}(p)
\bar d(p)] ,
\end{equation}
where $N_{g}$ denotes the hermitean ghost number operator.

Our next step consists in constructing the BRST charge operator. One can
verify that
\begin{equation}
Q \equiv - \int_{0}^{\infty} dp [{\bar d}^{\dag}(p) b(p) + b^{\dag}(p)
\bar d(p)], \label{9}
\end{equation}
correctly implements the quantum analog of the global transformation
(\ref{10}). Furthermore, $ Q^{2} = 0$. By using the commutation relations
(\ref{two}) and (\ref{8}) one arrives to the quartet algebra \cite{Ku}
\begin{mathletters}
\label{qu}
\begin{eqnarray}
[Q\,,\,a(p)]& = & i \bar d(p), \\
\phantom a [Q\,,\,b(p)] & = & 0 , \\
\phantom a \{Q\,,\bar d(p)\} & = & 0 , \\
\phantom a \{Q\,, d(p)\} & = & - b(p) .
\end{eqnarray}
\end{mathletters}
We now recall that physical states are required to verify $ Q \vert {\it
phys}> = 0 $. Hence, by the quartet mechanism \cite{Ku} , all physical
states , with the exception of the vacuum, are zero--norm states. The
physical S--matrix is just the identity operator and  $<~0~\vert~H_{0} +
H_{0}^{g} \vert 0>  = 0 $.

Thus, the addition of ghosts render the theory consistent but, however,
trivial. We then conclude that linear constraints do not provide
an efficient mechanism to generate chiral bosons.
We mention that several models for chiral bosons not based on the linear
constraint, and therefore free of the above difficulties, have been proposed
in the past \cite{Si,La,Ja,Gi}.
\acknowledgements

This work has been supported in part by Conselho Nacional de Desenvolvimento
Cient\'{\i}fico e Tecnol\'ogico (CNPq), Brazil, and Funda\c{c}\~ao
de Amparo \`a Pesquisa do Estado de S\~ao Paulo (FAPESP), Brazil.


\begin{references}
\bibitem{um} In this paper we use the Minkowski metric  $g^{00}=-g^{11}=1$,
$g^{\mu \nu}=0$ if $\mu \neq \nu$. Also $\epsilon^{\mu \nu} =-\epsilon^{\nu
\mu}$, $\epsilon^{01}=1$.
\end{references}
\end{document}